\documentclass[12pt]{article}
\usepackage{graphicx,amssymb,amsfonts,amsmath,epsfig}

\makeatletter
\DeclareSymbolFont{AMSb}{U}{msb}{m}{n}
\DeclareSymbolFontAlphabet{\mathbb}{AMSb}
\renewcommand{\section}{\@startsection{section}{1}{\z@}%
                                    {-7ex \@plus -1ex \@minus -.2ex}%
                                    {2.5ex \@plus.2ex}%
                                    {\normalfont\large\scshape\centering}}
\renewcommand{\subsection}{\@startsection{subsection}{2}{\z@}%
                                       {-5ex \@plus -1ex \@minus -.2ex}%
                                       {1.5ex \@plus.2ex}%
                                       {\normalfont\normalsize\scshape}}
\renewcommand{\subsubsection}{\@startsection{subsubsection}{3}{\z@}%
                                       {-5ex \@plus -1ex \@minus -.2ex}%
                                       {1.5ex \@plus.2ex}%
                                       {\normalfont\normalsize\scshape}}

\renewcommand\@seccntformat[1]{\ignorespaces\csname #1name\endcsname\space
                               \csname the#1\endcsname.\quad}   
\newdimen\captionmargin
\newdimen\captionindent
\newdimen\captionwidth
\newcommand{\captionfont}{\slshape}
\newcommand\@captionlabel[1]{\textsc{#1:}\space}
\long\def\@makecaption#1#2{%
  \vskip\abovecaptionskip
  \captionwidth\hsize
  \advance\captionwidth -2\captionmargin
  \sbox\@tempboxa{\@captionlabel{#1}\captionfont #2}%
  \ifdim \wd\@tempboxa >\captionwidth
    \ifdim\captionindent>\z@
      \advance\captionwidth -\captionindent
      \hskip\captionindent
    \fi
    \hskip\captionmargin
    \parbox[t]{\captionwidth}{\leavevmode\hskip-\captionindent
      \@captionlabel{#1}\captionfont #2}%
  \else
    \global \@minipagefalse
    \hb@xt@\hsize{\hfil\box\@tempboxa\hfil}%
  \fi
  \vskip\belowcaptionskip}
\def\eqnarray{%
   \stepcounter{equation}%
   \def\@currentlabel{\p@equation\theequation}%
   \global\@eqnswtrue
   \m@th
   \global\@eqcnt\z@
   \tabskip\@centering
   \let\\\@eqncr
   $$\everycr{}\halign to\displaywidth\bgroup
       \hskip\@centering$\displaystyle\tabskip\z@skip{##}$\@eqnsel
      &\global\@eqcnt\@ne$\;\hfil{##}$\hfil
      &\global\@eqcnt\tw@$\;\displaystyle{##}$\hfil\tabskip\@centering
      &\global\@eqcnt\thr@@ \hb@xt@\z@\bgroup\hss##\egroup
         \tabskip\z@skip
      \cr}
\ifcase \@ptsize
\or
\or
\fi
  \divide\@tempdima\baselineskip
  \@tempcnta=\@tempdima
\makeatother
\begin{document}

%
%

\renewcommand{\theequation}{\arabic{section}.\arabic{equation}}
\renewcommand{\thefigure}{\arabic{figure}}
\newcommand{\gapprox}{%
\mathrel{%
\setbox0=\hbox{$>$}\raise0.6ex\copy0\kern-\wd0\lower0.65ex\hbox{$\sim$}}}
\textwidth 165mm \textheight 220mm \topmargin 0pt \oddsidemargin 2mm
\def\ib{{\bar \imath}}
\def\jb{{\bar \jmath}}

\newcommand{\ft}[2]{{\textstyle\frac{#1}{#2}}}
\newcommand{\be}{\begin{equation}}
\newcommand{\ee}{\end{equation}}
\newcommand{\bea}{\begin{eqnarray}}
\newcommand{\eea}{\end{eqnarray}}
\newcommand{\Identity}{{1\!\rm l}}
\newcommand{\cx}{\overset{\circ}{x}_2}
\def\CN{$\mathcal{N}$}
\def\CH{$\mathcal{H}$}
\def\hg{\hat{g}}
\newcommand{\bref}[1]{(\ref{#1})}
\def\espai{\;\;\;\;\;\;}
\def\zespai{\;\;\;\;}
\def\avall{\vspace{0.5cm}}
\newtheorem{theorem}{Theorem}
\newtheorem{acknowledgement}{Acknowledgment}
\newtheorem{algorithm}{Algorithm}
\newtheorem{axiom}{Axiom}
\newtheorem{case}{Case}
\newtheorem{claim}{Claim}
\newtheorem{conclusion}{Conclusion}
\newtheorem{condition}{Condition}
\newtheorem{conjecture}{Conjecture}
\newtheorem{corollary}{Corollary}
\newtheorem{criterion}{Criterion}
\newtheorem{defi}{Definition}
\newtheorem{example}{Example}
\newtheorem{exercise}{Exercise}
\newtheorem{lemma}{Lemma}
\newtheorem{notation}{Notation}
\newtheorem{problem}{Problem}
\newtheorem{prop}{Proposition}
\newtheorem{rem}{{\it Remark}}
\newtheorem{solution}{Solution}
\newtheorem{summary}{Summary}
\numberwithin{equation}{section}
\newenvironment{pf}[1][Proof]{\noindent{\it {#1.}} }{\ \rule{0.5em}{0.5em}}
\newenvironment{ex}[1][Example]{\noindent{\it {#1.}}}

\thispagestyle{empty}


\begin{center}

{\LARGE\scshape Back-reaction as a Quantum correction
\par}
\vskip15mm

\textsc{Oscar Lorente-Esp\'{i}n}
\par\bigskip
{\em
Departament de F{\'\i}sica i Enginyeria Nuclear,
Universitat Polit\`ecnica de Catalunya,\\
Comte Urgell, 187, E-08036 Barcelona, Spain.}\\[.1cm]
\vspace{5mm}
\end{center}

\section*{Abstract}
In this work we will show how the back-reaction can be treated as a quantum correction. 
The novel semi-classical approach presented here consists in the introduction 
of adequate quantum corrections into the $r-t$ sector of the black hole metric. Thus, we will obtain corrected values for 
the temperature, entropy and emission rate, which at leading order coincide with the results in the tunneling frame.
We have also applied this technique to the 
Little String Theory. Interestingly, we have found similar results for the entropy as using string one-loop calculations. 

\vspace{3mm} \vfill{ \hrule width 5.cm \vskip 2.mm {\small
\noindent E-mail: oscar.lorente-espin@upc.edu }}

\newpage
\setcounter{page}{1}


\tableofcontents       %
\vskip 1cm             %

\setcounter{equation}{0}

\section{Introduction}
Since the pioneering proposal of Hawking that black holes can radiate \cite{Hawking:1974sw}, much work has been done in 
order to obtain a complete theory of quantum gravity. In the work of Hawking a paradox had emerged; the information loss 
paradox with the apparent violation of unitarity principle has consequences on well-established quantum mechanics. 
A recent effort has been done in order to get more insight on the process of the radiation emission by black holes. Using 
different semi-classical approaches such as: the tunneling method \cite{Parikh:1999mf}, the complex path analysis 
\cite{Srinivasan:1998ty}, or the cancellation of gravitational anomalies \cite{Robinson:2005pd}; we are able to analyze the 
back-reaction of the metric.\par
During the radiation emission of a black hole we enforce energy conservation,
thus the metric back-reacts and the event horizon shrinks. 
When the black hole radiates the total ADM mass is conserved, whereas the mass of the 
black hole decreases by the same amount of the energy that has been released by the black hole by emission. 
According to the heuristic picture most commonly considered \cite{Hawking:1977wz}, the quantum vacuum fluctuations 
generate a pair of virtual particles; one member of the pair, for example the anti-particle, falls down to the black hole 
while the other member of the pair, i.e. the particle, escapes towards the asymptotic infinity. The net effect would be as 
if the black hole had emitted a particle at the expenses of slowly decreasing its 
mass. Accordingly, we must consider the quantum nature of the 
emission process; thereby, we have been led to introduce quantum perturbations into the original static metric of 
the black hole in order to evaluate the back-reaction.\par 
In this work we have considered a perturbed general metric with some sort of perturbations with quantum character.
Eventually, we have the main aim to show that the back-reaction of the metric, i.e. imposing energy conservation, can be 
viewed as a quantum perturbation. 
Furthermore, we have analyze the introduction of the same sort of perturbations into the LST, a theory that is claimed to be 
dual to a certain string theory. Likewise we have been leading to similar conclusions previously studied in the literature, 
but using in this work a novel semi-classical approach. 

\section{Quantum correction on the metric}
Consider a general metric in conformal-string frame with spherical symmetry defined in a $d$-dimensional space-time,
\begin{equation}
 \label{generalmetric}
ds^2=-f(r)dt^2+\frac{g(r)}{f(r)}dr^2+h(r)r^2d\Omega_{d-2}^2 \;.
\end{equation}
The event horizon is found at the radial coordinate position $r_{0}$ and $d\Omega_{d-2}^2$ defines the $(d-2)$-sphere. 
Since the radiation emission depends only on the $r-t$ sector of the metric, we are going to slightly modify those 
terms of the metric, furthermore we want that these changes on the metric accounts for quantum effects. 
In \cite{Banerjee:2008cf} the authors introduced quantum corrections considering all the terms in the expansion of 
a single particle action. Therefore, motivated by this work, we introduce the following perturbations on the radial and 
time part of the metric (\ref{generalmetric}),
\begin{equation}
 \label{perturbations}
\delta g_{tt}=-f(r)\sum_{i}\frac{\xi_{i}\hbar^i}{\xi_{i}\hbar^i+r_{0}^{(d-2)i}} \;\;\;,\;\;\; 
\delta g_{rr}=\frac{g(r)}{f(r)}\sum_{i}\xi_{i}\frac{\hbar^i}{r_{0}^{(d-2)i}}\;,
\end{equation}
thus the slightly perturbed metric, $\hat{g}_{\mu\nu}=g^{(0)}_{\mu\nu}+\delta g_{\mu\nu}$, can be written as
\begin{equation}
 \label{perturbedmetric}
\hat{ds}^2=-f(r)\left(1+\sum_{i}\xi_{i}\frac{\hbar^i}{r_{0}^{(d-2)i}}\right)^{-1}dt^2+
\frac{g(r)}{f(r)}\left(1+\sum_{i}\xi_{i}\frac{\hbar^i}{r_{0}^{(d-2)i}}\right)dr^2+h(r)r^2d\Omega^2_{d-2} \;,
\end{equation}
where $\xi_{i}$ are positive definite dimensionless parameters. This choice of the perturbations has been motivated by 
dimensional analysis. Noticing that the 
reduced Planck length $(\tilde{l}_{P}=\frac{{l}_{P}}{2\pi})$ in a $d$-dimensional space-time is defined as 
$\tilde{l}_{P}^{d-2}=\frac{\hbar G^{(d)}}{c^{3}}$, where $G^{(d)}$ is the $d$-dimensional Newton's constant. 
In natural units $(G=c=1)$ we obtain the following dimensional relation 
$[\tilde{l}_{P}^{d-2}]=[\hbar]$. Since for the black hole metric (\ref{generalmetric}) we have only one parameter with 
length dimensions, i.e. the event horizon $r_{0}$; we conclude that $r_{0}^{d-2}$ must be proportional to $\hbar$.\par
The perturbed metric expression (\ref{perturbedmetric}) deserves a few comments. Firstly, we should verify whether it is 
a solution of the Einstein equations. In fact, we notice that it is the case since the perturbations are independent 
of any of the coordinates. Secondly, we point out the modification of the particles velocity in the region near the 
event horizon. 
Causal propagation is limited to time-like and null particle trajectories with respect to the background 
(\ref{perturbedmetric}), therefore in the case of null coordinates we find that the maximum velocity of photons has been 
shifted to a new value, 
\begin{equation}
 \label{velocityshift}
\hat{c}=c\left(1+\sum_{i}\xi_{i}\frac{\hbar^i}{r_{0}^{(d-2)i}}\right)^{-1} \;.
\end{equation}
In any case we do not obtain superluminal propagation velocities.
Eventually, we verify that the null energy condition is not affected by the inclusion of quantum perturbations, thus 
$T_{\mu\nu}e^{\mu}e^{\nu}\geq0$, or equivalently  $R_{\mu\nu}e^{\mu}e^{\nu}\geq0$ for any null vector $e^{\mu}$, is 
accomplished near the event horizon.\par 
Next, we are interested in to study how the Hawking temperature of the 
black hole is modified by the above perturbations. Then, if we 
introduce the euclidean time, 
$\tau=it$, we get the corresponding Euclidean positive definite metric. Furthermore, taking into account the definition 
of the proper length, $d\rho^2=g_{rr}dr^2$, together with the expansion of the metric function near the event horizon, 
$f(r)=f'(r_{0})(r-r_{0})$, then we can define a new radial coordinate as 
$\rho=2\sqrt{\frac{g(r)(r-r_{0})}{f'(r)}}\big|_{r\rightarrow r_{0}}
\left(1+\sum_{i}\xi_{i}\frac{\hbar^i}{r_{0}^{(d-2)i}}\right)^{1/2}$.
Hence we write the metric in Rindler coordinates,
\begin{equation}
 \label{euclideanmetric}
\hat{ds}^{2}_{E}=\rho^2\left(\frac{f'(r)}{2\sqrt{g(r)}}\big|_{r\rightarrow r_{0}}
\left(1+\sum_{i}\xi_{i}\frac{\hbar^i}{r_{0}^{(d-2)i}}\right)^{-1}d\tau\right)^2+d\rho^2+h(r)r^2d\Omega^2_{d-2} \;,
\end{equation}
where we point out the presence of the surface gravity modified by the correction terms, 
\begin{equation}
 \label{surfacegravity}
\hat{\kappa}=\frac{f'(r)}{2\sqrt{g(r)}}\big|_{r\rightarrow r_{0}}
\left(1+\sum_{i}\xi_{i}\frac{\hbar^i}{r_{0}^{(d-2)i}}\right)^{-1} \;.
\end{equation}
We can remove the apparent conical singularity at the event horizon in (\ref{euclideanmetric}) by identifying the 
imaginary (Euclidean) time coordinate with the period $\beta=\frac{2\pi}{\hat{\kappa}}$. Thereby we find that the 
effective temperature corresponding to the perturbed black hole is
\begin{equation}
 \label{temp1}
\hat{T}=\frac{\hbar\hat{\kappa}}{2\pi} \;.
\end{equation}
In this equation we can observe that the new temperature is just the standard Hawking temperature
\begin{equation}
 \label{temp0}
T_{H}=\frac{\hbar}{4\pi}\frac{f'(r)}{\sqrt{g(r)}}\big|_{r\rightarrow r_{0}}\;,
\end{equation}
corrected by quantum perturbations.

\section{Back-reaction viewed as a quantum correction}
In the spirit of the above section we would like to analyze how the metric is affected by the back-reaction, and 
consequently 
if we can consider such back-reaction of the metric as a quantum effect. When the black hole emits a particle 
with energy $\omega$, the metric back-reacts in a quantity proportional to the energy released by the black hole. The total 
ADM energy is conserved but the mass of the black hole is modified to $M\rightarrow M-\omega$, thereby the event horizon 
shrinks leading to the tunnel emission of the particle. Eventually, energy conservation 
leading us to non-thermal emission spectrum of the black holes, as one can see in \cite{Parikh:1999mf} and 
\cite{LorenteEspin:2011pu}. Motivated by the idea that the emitted particles 
are quantum fields whose energy, $\omega$ in natural units ($\hbar=1$), is also quantized; our aim is to show if we can 
treat the back-reaction of the metric as a quantum perturbation.\par
In order to interpret properly the quantum perturbation of the back-reacted metric, it is useful to show the relation 
between the mass and the event horizon of the black hole. For that purpose we have calculated the Komar integral 
\footnote{We will consider static backgrounds whose components of the metric are time-independent at infinity, 
therefore the Komar energy is equivalent to the ADM mass: 
$M=\frac{1}{8(d-3)\pi G^{(d)}}\int_{\partial\Sigma}d^{(d-2)}x\sqrt{g^{(d-2)}}\;e_{\mu}e_{\nu}\nabla^{\mu}K^{\nu}$. The 
integral is taken over the boundary of an hypersurface $\Sigma$, and $e_{\mu}$ are the vielbeins.} 
associated with the time-like Killing vector $K^{\nu}$. For 
the general background (\ref{generalmetric}) we have found the following relation,
\begin{equation}
 \label{admmass}
M=\frac{{\rm Vol}({\bf S}^{d-2})}{8(d-3)\pi G^{(d)}}\;\frac{f'(r)}{\sqrt{g(r)}}\;
\left(r\sqrt{h(r)}\right)^{d-2}\big|_{r\rightarrow r_{0}} \;,
\end{equation}
where ${\rm Vol}({\bf S}^{d-2})$ stands for the volume of the $(d-2)-$sphere,
prime denotes derivative with respect to radial coordinate, and all quantities are evaluated at the event horizon.
Moreover, we also impose the following three conditions on the space-time metric:
\begin{enumerate}
 \item Spherical symmetry.
 \item The background is asymptotically flat Minkowski. 
 \item The metric function $f(r)$ is expressed as $f(r)=1-\left(\frac{r_{0}}{r}\right)^{d-3}$, depending on the mass 
       through the event 
       horizon $r_{0}$. For future convenience we write the metric functions $g(r)$ and $h(r)$ as 
       $\left(1+\frac{r_{i,j}^2}{r^2}\right)$, depending on the charges $r_{i}$ and $r_{j}$ which are different from the 
       mass charge. With this choice for the metric functions we see from the relation (\ref{admmass}) that 
       $M\propto r_{0}^{d-3}$.
\end{enumerate}
Taking into account the above three conditions, and expanding in the energy of the emitted particle $\omega$, 
we eventually write (\ref{generalmetric}) as
\begin{equation}
 \label{backreactedmetric}
\tilde{ds}^2=-\tilde{f}(r)dt^2+\frac{g(r)}{\tilde{f}(r)}dr^2+h(r)r^2d\Omega^2_{d-2} \;,
\end{equation}
where we have defined the new metric function $f(r)$ as
\begin{equation}
 \label{backf}
\tilde{f}(r)=f(r)+\frac{1}{r^{d-3}}\sum_{i}\frac{\omega^i}{r_{0}^{(d-3)(i-1)}} \;.
\end{equation}
We motivate this expression for the expansion in the energy $\omega$ of the particle on dimensional analysis, since we 
have just seen that $r_{0}^{d-3}$ has energy-mass dimension. 
Proceeding as in the above section we find the effective temperature, which is 
$\tilde{T}=\frac{\hbar}{4\pi}\frac{\tilde{f'}(r)}{\sqrt{g(r)}}\big|_{r\rightarrow r_{0}}$. Thus
taking the derivative of (\ref{backf}) at the event horizon, we eventually obtain for the temperature corresponding 
to the back-reacted metric,
\begin{equation}
 \label{temp3}
\tilde{T}=T_{H}-\frac{\hbar(d-3)}{4\pi r_{0}^{d-2}\sqrt{g(r_{0})}}\sum_{i}\frac{\omega^i}{r_{0}^{(d-3)(i-1)}} \;.
\end{equation}
Since the heat capacity is negative, we can verify that this expression for the temperature works properly increasing its 
value when the black hole emits a 
particle of energy $\omega$. To see this, we can rewrite equation (\ref{temp3}) using the definition of the Hawking 
temperature (\ref{temp0}) together with the above third condition, hence we get for the effective temperature: 
$\tilde{T}=\frac{\hbar(d-3)}{4\pi\sqrt{g(r_{0})}}\left(\frac{1}{r_{0}}-\frac{1}{r_{0}^{d-2}}
\sum\frac{\omega^i}{r_{0}^{(d-3)(i-1)}}\right)$. Since the black hole shrinks its event horizon in a quantity proportional 
to $\omega^{1/(d-3)}$ when emits a particle, we see from this last expression that at low energies the 
temperature increases with respect to the standard Hawking temperature (\ref{temp0}).\par
Finally, if we compare the two expressions for the temperatures (\ref{temp1}) and (\ref{temp3}), we obtain definite values 
for the dimensionless parameters in terms of the released energy,
\begin{equation}
 \label{parameters}
\xi_{i}=\left(\frac{r_{0}^{d-2}}{\hbar}\right)^{i}\frac{\omega^i}{r_{0}^{(d-3)i}-\omega^i} \;.
\end{equation}
Therefore, looking at the metric (\ref{perturbedmetric}) and its corresponding temperature (\ref{temp1}), we conclude 
that back-reaction can be treated as a quantum perturbation leading us to the following expressions for the perturbed 
metric and effective temperature respectively,
\begin{equation}
 \label{perturbedmetric1}
\hat{ds}^2=-f(r)\left(1+\sum_{i}\frac{\omega^i}{r_{0}^{(d-3)i}-\omega^i}\right)^{-1}dt^2+
\frac{g(r)}{f(r)}\left(1+\sum_{i}\frac{\omega^i}{r_{0}^{(d-3)i}-\omega^i}\right)dr^2+h(r)r^2d\Omega^2_{d-2} \;,
\end{equation}
\begin{equation}
 \label{effectivetemp1}
\hat{T}=T_{H}\left(1+\sum_{i}\frac{\omega^i}{r_{0}^{(d-3)i}-\omega^i}\right)^{-1} \;.
\end{equation}

We are going to specify all what have been said in a simple four-dimensional, static and spherically symmetric background. 
Thereby we consider a Schwarzschild black hole which is asymptotically flat Minkowski, the metric functions are defined as: 
$f(r)=1-\frac{2M}{r}$, $g(r)=h(r)=1$ and the event horizon is at $r_{0}=2M$ in natural units. From (\ref{perturbedmetric1}) 
we write the perturbed back-reacted metric as,
\begin{equation}
 \label{perturbedmetricschwarz}
\hat{ds}^2=-\left(1-\frac{r_{0}}{r}\right)\left(1+\sum_{i}\frac{\omega^i}{r_{0}^{i}-\omega^i}\right)^{-1}dt^2+
\frac{1}{\left(1-\frac{r_{0}}{r}\right)}\left(1+\sum_{i}\frac{\omega^i}{r_{0}^{i}-\omega^i}\right)dr^2+
r^2d\Omega^2_{2} \;.
\end{equation}
The Hawking temperature corresponding to a Schwarzschild black hole is $T_{H}=\frac{1}{8\pi M}$. Then when the black 
hole emits a single particle with energy $\omega$, the new effective temperature at first order in energy expansion will be
\begin{equation}
 \label{tempschwarz}
\hat{T}=\frac{1}{8\pi(M-\omega)}\left(1+\frac{\omega}{2M-\omega}\right)^{-1} \;.
\end{equation}
Likewise at semi-classical level we calculate the Bekenstein-Hawking entropy using the area law, $S_{BH}=\frac{A}{4}$, 
in the presence of back-reaction effects, 
\begin{equation}
 \label{entropyareaschwarz}
\hat{S}_{BH}=4\pi (M-\omega)^2 \;.
\end{equation}
Furthermore we also calculate the emission rate through the relation $\Gamma\propto e^{-\omega/T}$, \cite{Hartle:1976tp}. 
Therefore using the effective temperature (\ref{tempschwarz}) we obtain
\begin{equation}
 \label{rateschwarz}
\Gamma\propto e^{-8\pi\omega(M-\omega)\left(1+\frac{\omega}{2M-\omega}\right)} \;.
\end{equation}
At low energies we notice that the emission rate can be written semi-classically as 
\begin{equation}
 \label{rateschwarz1}
\Gamma\propto e^{\Delta \hat{S}_{BH}} \;,
\end{equation}
being the initial entropy $S_{BH}^{(0)}=4\pi M^{2}$, and the final entropy (\ref{entropyareaschwarz}). We point out that 
this result coincides with the results in \cite{Parikh:1999mf}, where the emission rate also matches the statistical 
mechanics picture. We see that deviation from thermal behavior when a black hole 
emits is due to the energy conservation, moreover the temperature increases while the entropy of the black hole decreases 
properly during 
the emission process. Thus summarizing, we have seen that all the semi-classical results concerning the emission of 
particles are recovered when we consider the back-reaction as a quantum correction.\par
On the other hand, instead of calculating the entropy using the area law, we can evaluate the entropy through the first law 
of thermodynamics: $dM=TdS$, in the presence of back-reaction effects. Thus using (\ref{tempschwarz}) for the temperature 
we obtain
\begin{equation}
 \label{entropy1lawschwarz}
\hat{S}=4\pi M^2-4\pi M\omega-2\pi\omega^2 {\rm log}(2M-\omega)+\pi\omega^2+{\cal O}(\omega^{3}) \;.
\end{equation}
The first term is just the semi-classical area law, $S_{BH}=\frac{A}{4}$. Nevertheless, we interestingly point out the 
presence of a logarithmic correction term, which is considered a one-loop correction term over the classical thermodynamics. 

\section{An example in string theory}
In this section we are going to illustrate the preceding techniques in a stringy black hole background, moreover we will 
elucidate some thermodynamical aspects. We consider a stack of $N$ coincident NS5-branes in type II string theory in the 
limit of a vanishing asymptotic value for the string coupling, $g_{s}\rightarrow0$, keeping the energy density 
$\frac{E}{m_{s}}={\rm fixed}$, where $m_{s}$ is the string mass. In 
this near-horizon limit the theory becomes free in the bulk but strongly interacting on the brane, thus defining a 
non-gravitational, 
six-dimensional non-local field theory. This theory is known as Little String Theory (LST) and is claimed to be dual to 
a string theory background, see \cite{Kutasov:2001uf,Kutasov:2000jp} for a review.
The throat geometry corresponding to $N$ coincident non-extremal NS5-branes in the string frame is 
\begin{equation}
 \label{metriclst}
ds^{2}=-f(r)dt^2+\frac{A(r)}{f(r)}dr^2+A(r)r^2d\Omega_{3}^2+\sum_{j=1}^{5}dx_{j}^2\;,
\end{equation}
where the coordinates $x_{j}$ corresponds to flat spatial directions along the 5-branes. 
The metric functions are defined as
\begin{equation}
 \label{metricf}
f(r)=1-\frac{r_{0}^2}{r^2}\;\;\;,\;\;\;A(r)=\chi+\frac{N}{m_{s}^2r^2}\;,
\end{equation}
where the location of the event horizon corresponds to $r=r_{0}$. We define the parameter $\chi$ which takes the values
1 for NS5 model and 0 for LST, since only for this values of $\chi$ exist a supergravity solution. According to the 
holographic principle, the high energy spectrum of this dual string theory should be 
approximated by certain black hole in the background (\ref{metriclst}), whose boundary near horizon geometry is 
$R^{5,1}\times R\times S^{3}$.\par 
The Hawking temperature is
\begin{equation}
 \label{templst}
T_{H}=\frac{\hbar}{2\pi\sqrt{\chi r_{0}^2+\frac{N}{m_{s}^2}}}\;.
\end{equation}
Following the above techniques we introduce the perturbations at first order modifying the function metric $f(r)$ by 
\begin{equation}
 \label{fperturbed}
f(r)\rightarrow f(r)\left(1+\frac{\omega}{r_{0}^2-\omega}\right)^{-1} \;.
\end{equation}
Now the effective temperature will be 
\begin{equation}
 \label{templst1}
\hat{T}=T_{H}\left(1+\frac{\omega}{r_{0}^2-\omega}\right)^{-1} \;,
\end{equation}
and also from (\ref{admmass}) we have calculated the ADM mass corresponding to the NS5 and LST black hole,
\begin{equation}
 \label{masslst}
M=\frac{{\rm Vol}({\bf R}^5)\pi}{4G^{(10)}}\left(\chi r_{0}^2+\frac{N}{m_{s}^2}\right) \;,
\end{equation}
where ${\rm Vol}({\bf R}^5)$ stands for the volume of the NS5-branes. In \cite{Cotrone:2007qa} and previously in 
\cite{Kutasov:2000jp} it was found that the 
Helmholtz free energy vanishes, ${\cal F}=E-TS=0$. Thereby the entropy coincides with the semi-classical area law entropy,
\begin{equation}
 \label{entropylst}
S_{BH}=\frac{A}{4G^{(10)}\hbar}=\frac{{\rm Vol}({\bf R}^5)\pi^2}{2G^{(10)}\hbar}
\left(\chi r_{0}^2+\frac{N}{m_{s}^2}\right)^{3/2} \;.
\end{equation}
Then, for a emission process taking into account back-reaction we shall consider the effective temperature (\ref{templst1}). 
Moreover, considering the relation between the mass and the event horizon (\ref{masslst}), we calculate the emission rate 
$\Gamma\propto e^{\omega/\hat{T}}$ at low energies,
\begin{equation}
 \label{ratelst}
\Gamma\propto e^{-\frac{\omega}{T_{H}}\left(1+\frac{\omega}{2M}+{\cal O}(\omega^2)\right)} \;,
\end{equation}
which is in accordance with the statistical mechanics result, $\Gamma\propto e^{\Delta \hat{S}_{BH}}$, being 
$\hat{S}_{BH}\propto(M-\omega)^{3/2}$ the Bekenstein-Hawking entropy after the emission. 
This last result coincides entirely with the result in \cite{LorenteEspin:2011pu} for the NS5 black hole, however it is 
not the case for LST which disagrees in the factor $e^{\sim\frac{\omega}{2M}}$. In that work we calculated the emission 
rate modifying naively the mass factor that appears in the temperature, i.e. $M\rightarrow M-\omega$, without 
taking into account any temperature correction factor. Meanwhile in this work we have considered the effective temperature 
(\ref{templst1}), obtaining in this way an interesting deviation from pure thermal behavior found in 
\cite{LorenteEspin:2011pu,LorenteEspin:2007gz}. Therefore this result signals some sort of correction over the classical 
thermodynamics.

\subsection{Discussion}
The thermodynamics of the near horizon limit of NS5 presents a Hagedorn behavior, where the statistical mechanics of any 
string theory breaks-down. At very high energy density, one can see from (\ref{templst}) that the Hagedorn temperature of 
LST is independent of the 
mass, thus at leading order the thermodynamics will be completely degenerate with a constant temperature. Furthermore, the 
entropy will be proportional to the energy, $E=T_{H}S$, hence the free energy is expected to vanish. In 
\cite{Harmark:2000hw} the authors implemented string one-loop corrections in the near horizon limit of the  NS5-brane 
thermodynamics to 
explain the Hagedorn behavior of LST. That corrections expand the phase space and introduce small deviations from the 
constant Hagedorn temperature. On the other hand, in the high energy regime LST become weakly coupled, thus being able to 
perform 
a perturbative holographic analysis, see \cite{Kutasov:2000jp}. In this work it is shown that LST has a Hagedorn density 
of states that grows exponentially: $\rho=e^{S(E)}\sim E^{\alpha}e^{E/T_{H}}$, then the authors computed the genus one 
correction to both the temperature and the density of states. 
As in \cite{Harmark:2000hw} they found an entropy-energy relation with logarithmic corrections.\par
In our preceding study we have introduced, at semi-classical level, some sort of quantum energy corrections into the 
temperature 
for back-reaction processes. Now, our aim is to calculate the corrected entropy for the LST black hole. 
If we calculate the ADM mass-energy 
in Einstein frame, $ds_{E}^2=\sqrt{g_{s}e^{-\Phi}}ds^2$, we find the relation between the mass and the event horizon. Hence 
we can write the corrected Hawking temperature corresponding to LST as
\begin{equation}
 \label{templst2}
\hat{T}_{H}=\frac{\hbar}{2\pi\sqrt{\frac{N}{m_{s}^2}}}
\left(1+\frac{\omega}{\frac{4G^{(10)}M}{{\rm Vol}({\bf R}^5)\pi}-\omega}\right)^{-1} \;.
\end{equation}
We note that this temperature decreases when the LST black hole emits radiation, thereby the specific heat will be positive. 
Moreover, we verify that the Hagedorn temperature is the maximum temperature reached by the system and cannot be crossed. 
Now, integrating $dM/\hat{T}_{H}$ we obtain the corrected entropy,
\begin{equation}
 \label{entropylstcorrected}
\hat{S}(M)\approx\frac{M}{T_{H}}+\frac{{\rm Vol}({\bf R}^5)\pi^2\omega\sqrt{N}}{2G^{(10)}\hbar m_{s}}{\rm log}(M)+
{\cal O}\left(\frac{1}{M}\right) \;.
\end{equation}
Thus, we have obtained the classical Bekenstein-Hawking term plus logarithmic corrections to the entropy of LST. As in 
\cite{Kutasov:2000jp} and \cite{Harmark:2000hw} we have found that the logarithmic term  is 
${\rm Vol}({\bf R}^5)$-dependent.\par
Finally, we would like to make a last remark on the thermodynamics of LST. Looking at the temperature (\ref{templst2}), we 
are tempted to modify the plus sign of the correction factor by minus sign. The purpose of this change of sign is to fit 
the usual behavior of classical black holes, e.g. Schwarzschild-like black holes increase its temperature when they emit 
radiation, and have a negative specific heat. In this case we point out that, in accordance with \cite{Kutasov:2000jp}, 
the logarithmic correction term of the entropy (\ref{entropylstcorrected}) will be negative, the temperature 
(\ref{templst2}) will be above the Hagedorn temperature and the specific heat will be negative, therefore the 
thermodynamics will be unstable. Thus, if we perform a semi-classical analysis introducing quantum corrections on the 
metric, we are able to obtain similar results as working with string loop corrections.

\section{Conclusions}
In this work we have shown that back-reaction can be treated as a quantum perturbation on the metric. Using the novel 
semi-classical approach proposed here, we have reproduced 
properly results for the emission rate and the entropy, not only in the simple Schwarzschild case but as well for LST. When 
the function metric is modified with the appropriate corrections, one obtains a corrected temperature that lead us to the 
classical Bekenstein-Hawking entropy plus string one-loop logarithmic terms. Furthermore, we have seen that the 
temperature of LST, namely the Hagedorn temperature, decreases when we consider the quantum perturbations. 
Hereafter, we would be interested in to explore 
how the introduction of adequate corrections into the metric could act as a bridge between the semi-classical side to the 
quantum gravity side, shedding light into some problems that are waiting for a complete quantum gravity theory.   

\vskip10mm
\noindent{\bf Acknowledgments:}\\
I thank P. Talavera for fruitful discussions and insightful comments on this work.


\begin{thebibliography}{50}

\bibitem{Hawking:1974sw}
  S.~W.~Hawking,
  ``Particle Creation by Black Holes,''
  Commun.\ Math.\ Phys.\  {\bf 43 } (1975)  199-220.

\bibitem{Parikh:1999mf}
  M.~K.~Parikh, F.~Wilczek,
  ``Hawking radiation as tunneling,''
  Phys.\ Rev.\ Lett.\  {\bf 85 } (2000)  5042-5045.
  [hep-th/9907001].

\bibitem{Srinivasan:1998ty}
  K.~Srinivasan, T.~Padmanabhan,
  ``Particle production and complex path analysis,''
  Phys.\ Rev.\  {\bf D60 } (1999)  024007.
  [gr-qc/9812028].

\bibitem{Robinson:2005pd}
  S.~P.~Robinson, F.~Wilczek,
  ``A Relationship between Hawking radiation and gravitational anomalies,''
  Phys.\ Rev.\ Lett.\  {\bf 95 } (2005)  011303.
  [gr-qc/0502074].

\bibitem{Hawking:1977wz}
  S.~W.~Hawking,
  ``The Quantum Mechanics of Black Holes,''
  Sci.\ Am.\  {\bf 236} (1977) 34.

\bibitem{Banerjee:2008cf}
  R.~Banerjee, B.~R.~Majhi,
  ``Quantum Tunneling Beyond Semiclassical Approximation,''
  JHEP {\bf 0806 } (2008)  095.
  [arXiv:0805.2220 [hep-th]].




 \bibitem{LorenteEspin:2011pu}
   O.~Lorente-Espin,
   ``Some considerations about NS5 and LST Hawking radiation,''
   Phys.\ Lett.\  {\bf B703 } (2011)  627-632.
   [arXiv:1107.0713 [hep-th]].

\bibitem{Hartle:1976tp}
  J.~B.~Hartle, S.~W.~Hawking,
  ``Path Integral Derivation of Black Hole Radiance,''
  Phys.\ Rev.\  {\bf D13 } (1976)  2188-2203.

\bibitem{Kutasov:2001uf}
  D.~Kutasov,
  ``Introduction to little string theory,''
  Lectures given at the Spring School on Superstrings and Related Matters. Trieste, 2-10 April 2001.


 \bibitem{Kutasov:2000jp}
   D.~Kutasov, D.~A.~Sahakyan,
   ``Comments on the thermodynamics of little string theory,''
   JHEP {\bf 0102 } (2001)  021.
   [hep-th/0012258].

\bibitem{Cotrone:2007qa}
  A.~L.~Cotrone, J.~M.~Pons, P.~Talavera,
  ``Notes on a SQCD-like plasma dual and holographic renormalization,''
  JHEP {\bf 0711 } (2007)  034.
  [arXiv:0706.2766 [hep-th]].

 \bibitem{LorenteEspin:2007gz}
   O.~Lorente-Espin, P.~Talavera,
   ``A Silence black hole: Hawking radiation at the Hagedorn temperature,''
   JHEP {\bf 0804 } (2008)  080.
   [arXiv:0710.3833 [hep-th]].

\bibitem{Harmark:2000hw}
  T.~Harmark and N.~A.~Obers,
  ``Hagedorn behavior of little string theory from string corrections to NS5-branes,''
  Phys.\ Lett.\ B {\bf 485} (2000) 285
  [hep-th/0005021].

\end{thebibliography}
\end{document}